\documentclass[aps,prl,twocolumn,showpacs,twoside]{revtex4}
\usepackage{graphicx}\usepackage{psfrag}
\newcommand{\mybf}[1]{{\bf #1}}
\renewcommand{\vec}{\mybf}
\begin{document}
\title{Aharonov-Bohm magnetization of mesoscopic rings caused by inelastic
relaxation.}
\author{O.L.Chalaev}
\affiliation{International School of Advanced Studies, via Beirut 4, 34014 Trieste, Italy}
\author{V.E.Kravtsov}
\affiliation{The Abdus Salam International Centre for Theoretical Physics,
P.O.B.
586, 34100 Trieste, Italy, \\ Landau Institute for Theoretical
Physics, 2 Kosygina st., 117940 Moscow, Russia.}
\begin{abstract}
The magnetization of a system of many mesoscopic rings under
non-equilibrium conditions is considered. The corresponding
disorder-averaged
current in a ring $I(\phi)$ is shown to be a sum of the `thermodynamic' and
`kinetic'
contributions both resulting from the electron-electron interaction. The
thermodynamic part can be expressed through the diagonal
matrix elements $J_{\mu\mu}$ of the current operator in the basis of exact many-body
eigenstates
and is a generalization of the equilibrium persistent current. The novel kinetic part
is present only out of equilibrium and is governed by the off-diagonal
matrix elements $J_{\mu\nu}$. It has drastically different temperature and magnetic
field behavior.

\end{abstract}
\maketitle
Recently there has been a considerable interest in mesoscopic phenomena under
non-equilibrium conditions. The relaxation of electron energy distribution function 
due to inelastic processes of electron interaction has been measured in mesoscopic
wires \cite{Pot}. The Kondo effect in quantum dots with finite bias has been
intensively discussed theoretically \cite{KNG,Hooly} and observed experimentally
\cite{Khov}.  

The non-equilibrium effects are also important for the magnetic
response of a system of mesoscopic rings \cite{KY,KA}.
In equilibrium a weak constant magnetic
field gives rise \cite{BIL}
to a magnetization of a system of many normal-metal mesoscopic rings that
corresponds to the average persistent current per ring
\begin{equation}
\label{PC} 
I^{(PC)}(\phi) =-\frac{\partial {\cal E}(\phi)}{\partial\phi}=
\sum_{n=1}^{\infty}I^{(PC)}_{n}\,\sin\left
(\frac{4\pi n
\phi}{\phi_{0}} \right),
\end{equation}
where $\phi$ is the magnetic flux through a ring, ${\cal E}(\phi)$ is the total
energy of interacting electrons in a ring averaged over the
disorder and the thermal ensembles, and    
$\phi_{0}=hc/e$ is the flux quantum.

The remarkable feature of the flux dependence in Eq.(\ref{PC}) is its periodicity and 
odd character.
The latter is due to the dissipationless nature of the equilibrium  persistent current. 
Indeed, in equilibrium
both directions of time are equivalent, so that time-reversal symmetry requires 
$I^{(PC)}(\phi)=-I^{(PC)}(-\phi)$. 
The periodic and odd in $\phi$ magnetization of the form Eq.(\ref{PC})
has been indeed observed \cite{Levy} in a sample containing $10^{7}$ mesoscopic
copper rings. 

However the 
{\it disorder-averaged} current in a
ring
must respect another symmetry related with the space homogeneity of the 
disorder-averaged system. This is space
reflection  about the ring diameter. It is easy to see that both the
current and 
the magnetic flux change
sign under such a reflection, so that the symmetry relationship 
$I(\phi)=-I(-\phi)$ should hold even in the case
where equilibrium and time reversal symmetry is not assumed. We thus arrive at the 
statement that the odd
character of $I(\phi)$ cannot be used as an evidence that the disorder-averaged 
current observed
in \cite{Levy} is an
{\it equilibrium} persistent current. 
The same is true for the $\phi_{0}/2$
periodicity. 

It has been shown by straightforward calculations \cite{KY,KA} that a {\it 
non-equilibrium}  dc current $I^{(dc)}(\phi)$ of
the
same form as
Eq.(\ref{PC}) indeed arises when the ring is driven out of equilibrium by an
external 
ac electric field. 

A non-equilibrium current is sustainable for a reasonably long time only if there
is an {\it external force}
acting on the
electron system. 
This force can produce a non-equilibrium dc current either 
by a {\it direct} action on electrons in the ring or indirectly. The case of
the external
ac electric field in Refs.\cite{KY,KA} can be considered as a  representative
example of
the direct 
effect. 

In the present Letter we consider another, indirect  mechanism of a non-equilibrium
current
which is related with the {\it relaxation} of
the {\it given non-equilibrium} electron
energy distribution created by an external force. 
The similar relaxation-induced mesoscopic photovoltaic effect has 
been considered in Ref.\cite{Spivak}. 
However, in contrast to Ref.\cite{Spivak}
we do not assume an
electron-phonon mechanism of relaxation which is extremely weak at low
temperatures \cite{KY1}.
Instead we consider much more effective inelastic processes due to {\it
electron-electron interaction}.

One can imagine the experimental geometry where the direct action of the
external
force on electrons in the ring is absent while the one-particle energy
distribution deviates from the Fermi-Dirac form. Consider a mesoscopic ring weakly
coupled
to the center of a wire connecting two reservoirs with the temperature
$T$ and the chemical potential difference $V$ (see Fig.1a) maintained by an
applied voltage. For a sufficiently short wire the distribution
of one-particle energies in the ring $f(E)$ is roughly a superposition of two
Fermi-Dirac distributions corresponding to two reservoirs \cite{Pot} (see Fig.1b).

\begin{figure}[h]
\psfrag{a}{c)}\psfrag{b}{b)}\psfrag{c}{a)}\psfrag{d}{a)}
\psfrag{I}{$I_E$}\psfrag{E}{$E$}\psfrag{n}{$\nu_E$}\psfrag{h}{$f(E)$}\psfrag{T}{$T$}
\psfrag{0}{$0$}
\psfrag{B}{$\mathbf{B}$}\psfrag{V}{$V$}\psfrag{e1}{$\scriptstyle{-V/2}$}\psfrag{e2}
{$\scriptstyle{V/2}$}
\includegraphics[width=3cm]{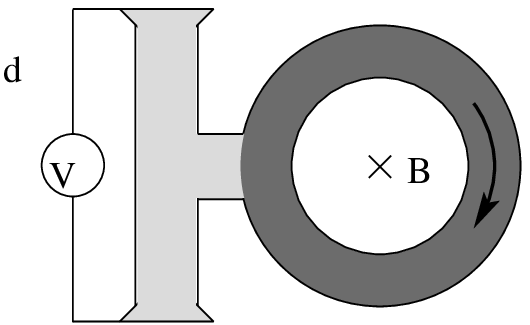}\hspace{1cm}
\includegraphics[width=3cm]
{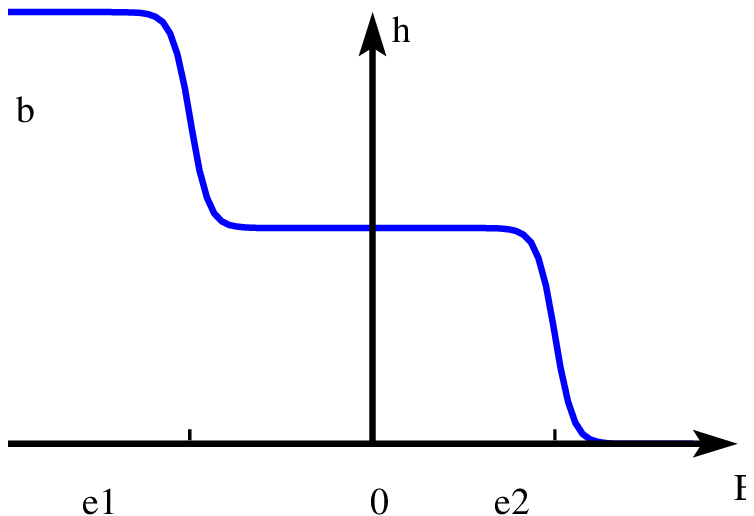}\hspace{1cm}
\includegraphics[width=3cm]{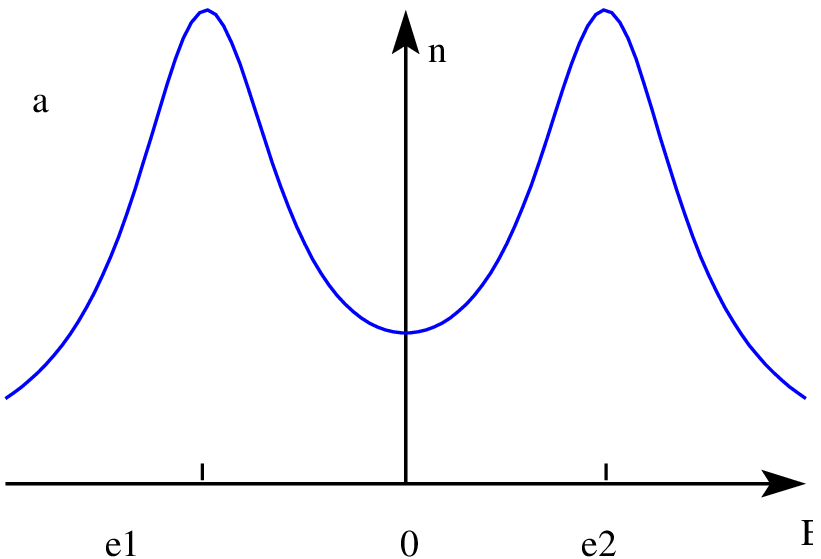}\\[.1cm]
\caption{a). The experimental geometry. b). Non-equilibrium
energy distribution function with two steps.
c). DoS for a Kondo system \cite{Khov}.}
\end{figure}

Below we identify the disorder-averaged dc
current $I^{(r)}(\phi)$ in a ring which is entirely due to the relaxation of this
distribution resulting from the {\it electron-electron} interaction. We show that
this relaxation-induced current can be represented in the
form Eq.(\ref{PC}) with the flux harmonics $I^{(r)}_{n}$ expressed through the
electron energy distribution function $f(E)=\frac{1}{2}(1-h_{E})$:
\begin{equation}
\label{fres}
I^{(r)}_{n}= C_{n}\frac{e}{h g}\,\int
dE dE'\,\left(\frac{\delta D_{E'}}{D_{0}} 
\right)\,\left(\frac{\partial R_{\omega}(E',E)}{\partial
\omega}\right)_{\omega=0},
\end{equation}
where  $-e$ is the electron charge, $g$ is the dimensionless conductance of the ring
with the perimeter $L$ and the cross-section area $S$,
$C_{n}=\frac{1}{(2\pi)^{2}}\sum_{m=1}^{\infty} \frac{1}{m^{2}}(1-e^{-2\pi n m})$,
and 
\begin{eqnarray}
\label{R}
R_{\omega}(E,E')&=&[(h_{E}
-h_{E-\omega})(1-h_{E'}h_{E'-\omega})\\ \nonumber &-& (h_{E'}
-h_{E'-\omega})(1-h_{E}h_{E-\omega})].
\end{eqnarray}
The diffusion coefficient $D_{E}=D_{0}+\delta D_{E}$ is supposed to have a small but
essentially energy-dependent part $\delta D_{E}\ll D_{0}$.
It is also assumed that the characteristic energy scale
of
the energy dependence in $\delta D_{E}$ is larger than the typical energy transfer
$\omega\sim E_{T}=\hbar D_{0}/L^{2}$.

The function $R_{\omega}(E,E')$ in Eq.(\ref{R}) is exactly the combination of the
electron energy distribution function $h_{E}$ that enters
in the inelastic collision integral \cite{R-S}:
\begin{equation}
\label{ke}
St(E)=\int dE' d\omega \,P(\omega)\,R_{\omega}(E,E').
\end{equation}
For the Fermi-Dirac distribution
$h_{E}=\tanh(E/2T)$ we have identically $R_{\omega}(E,E')=0$, and  both the
relaxation rate  Eq.(\ref{ke}) and the non-equilibrium current
Eq.(\ref{fres}) vanish. This
quantifies an intimate
relationship between them.

Eqs.(\ref{fres}),(\ref{R})  is the main result of the paper. It is valid
for the case of a pure potential disorder with no spin-orbit interaction
(orthogonal symmetry class) for $g\gg 1$ as long as $T\gg E_{T}$ and $n\ll
L_{\varphi}/L$, where $L_{\varphi}=(D_{E}\tau_{\varphi}^{({\rm
ring})})^{\frac{1}{2}}\gg 
L$ is
the
dephasing length and $\tau_{\varphi}^{({\rm ring})}$ is the
dephasing time in the
ring. 

The non-Fermi-Dirac form of $h_{E}$ is the necessary but not the sufficient
condition for $I^{(r)}_{n}$
to be non-zero. The global balance in Eq.(\ref{ke}) requires $\int
dE\,St(E)=0$ which results from the obvious identity $\int
dEdE'\,R_{\omega}(E,E')=0$. 
Then one concludes from Eq.(\ref{fres}) that $I^{(r)}_{n}=0$ unless
the electron diffusion coefficient $D_{E}=v^{2}\tau_{E}/3$  is
{\it
energy-dependent}.

For the white-noise impurity potential within the non-crossing
(self-consistent Born) approximation the product of the elastic scattering time
$\tau_{E}$ and the disorder-averaged one-electron density of states (DoS) $\nu_{E}$
is energy
independent  $\tau_{E}\nu_{E}=const$.
If
the energy dependence of $\nu_{E}$ is
stronger than the dependence of the
electron
velocity $v(E)$,  we obtain $D_{E}=D_{0}\nu/\nu_{E}$, 
where $\nu$ is the (one-spin) DoS  
outside the
region (near the Fermi energy) of the strong energy dependence.

Though the nature of the $E$-dependence of $\nu_{E}$ goes beyond
the scope of this paper we note that the 
electron-electron interaction
results in the strong energy dependence of the one-particle {\it tunnel} density of
states
$\nu^{({\rm tun})}_{E}$ exactly at the Fermi level \cite{AALee,LevShyt,Kamen}. In
our problem the
corresponding DoS $\nu^{(r)}_{E}$ should be
obtained from the non-perturbative treatment of the electron-electron interaction
and will be considered elsewhere. 
An alternative mechanism of the energy dependence (see Fig.1c) is the
Abrikosov-Suhl  
\cite{Abrikos} peaks in $\nu_{E}$ at  $E=\pm V/2$ that arise \cite{Khov}
because of the
Kondo effect.

Now we outline the derivation of Eq.(\ref{fres}). We start with the expression for
current density $J$ in terms of the components of the matrix Green's function
$\underline{G}=\left(\matrix{G^{R}& G^{K}\cr 0& G^{A}\cr}\right)$ in
the
Keldysh
technique \cite{Keld,R-S}
\begin{equation}
\label{init}
J= \frac{-i}{2}\,Tr\{\hat{J}G^{K}\},
\end{equation}
where $\hat{J}$ is the current density operator, $G^{R,A}$ are the retarded
(advanced) electron Green's functions, and $G^{K}$ is the Keldysh function.
The matrix Green's function $\underline{G}= \underline{G}_{0}+i  
\underline{G}_{0}\;\underline{\Sigma}^{F}\;\underline{G}_{0}+i
\underline{G}_{0}\;\underline{\Sigma}^{H}\;\underline{G}_{0}$ is calculated in
the first order in the
screened
electron interaction $\underline{U}=\left(\matrix{U^{R}& U^{K}\cr 0&
U^{A}\cr}\right)$,
where $\underline{G}_{0}$ is the matrix Green's function without electron
interaction in the presence of a static disorder potential and 
$\underline{\Sigma}^{F,H}$ are the Fock and the Hartree self-energy parts due to
electron interaction.

The electron energy distribution function $h_{E}$ enters in the theory
through the
Keldysh component
of the {\it unperturbed} matrix Green's function $\underline{G}_{0}$
via the ansatz \cite{Keld,R-S}
\begin{equation}
\label{ansatz}
G^{K}_{0}(E)= h_{E}\,[G^{R}_{0}(E)-G^{A}_{0}(E)].
\end{equation}
The effective interaction $\underline{U}$ is calculated in the random-phase
approximation which after averaging over disorder in the limit $g\gg 1$ yields:
\begin{equation}
\label{Uexp}
(U^{R}-U^{A})_{\omega,{\bf q}}= \int dE'\, U_{\omega,{\bf q}}(E')
\,(h_{E'}-h_{E'-\omega}),
\end{equation}
where
$2i\nu
D_{0}q^{2}\,U_{\omega,{\bf 
q}}(E')=(D_{0}^{2}q^{4}+\omega^{2})/(D_{E'}^{2}q^{4}+\omega^{2})$. 
It reduces to the well known expression \cite{R-S} in the
case of an energy-independent diffusion coefficient $\delta D_{E'}=0$ if one uses
the
identity $\int dE'\,(h_{E'}-h_{E'-\omega})=2\omega$. The Keldysh component of the
effective interaction $U^{K}_{\omega,{\bf q}}$ is given by Eq.(\ref{Uexp}) where
$(h_{E'}-h_{E'-\omega})$ is replaced by $(1-h_{E'}h_{E'-\omega})$. 

Rewriting the self-energy parts $\underline{\Sigma}^{F,H}$  explicitly
in
terms of the components $G_{0}^{R,A,K}$ and using the ansatz Eq.(\ref{ansatz})
one identifies \cite{KY1} three different interaction-induced contributions to the
current
Eq.(\ref{init}):
\begin{eqnarray}
\label{three}
{\bf I_{1}}&=&(A{\bf A R} - R {\bf A
R})\,[(h_{E}-h_{E-\omega})\,U_{\omega}^{K}-\\
\nonumber &-&(1-h_{E}h_{E-\omega})\,(U^{R}_{\omega}-U^{A}_{\omega})]\\
\nonumber
{\bf I_{2}}&=&A {\bf  R R}\, (U^{R}_{\omega}-pU^{R}_{0})\,(1-h_{E}h_{E-\omega})\\
\nonumber &+& R{\bf  A
A}\,(U^{A}_{\omega}-pU^{A}_{0})\,(1-h_{E}h_{E-\omega})
\\ \nonumber
{\bf I_{3}}&=&
(R{\bf RR}-A{\bf AA})h_E U^{K}_{\omega}-(1-h_{E}h_{E-\omega})\times\nonumber\\
\nonumber
& &[R{\bf RR}\,(U^{R}_{\omega}-p U^{R}_{0})+A{\bf AA}\,(U^{A}_{\omega}-p
U^{A}_{0})],
\end{eqnarray}  
where $A{\bf  A R}\equiv G^{A}_{0}(E-\omega)G^{{\bf A}}_{0}(E)\hat{J}G^{{\bf
R}}_{0}(E)$, and the integral over 
all $E$ and $\omega$ is assumed in Eq.(\ref{three}). The degeneracy factor 
$p=2$ for unpolarized electron spins and $p=1$ if spins are fully polarized by the
parallel magnetic field. 

The three contributions are very different in character. Since ${\bf I_{3}}$
contains only retarded or only advanced Green's functions it vanishes after
averaging over disorder at a constant chemical potential.  
The contribution ${\bf I_{2}}$ amounts to $I^{(PC)}_{n}$. In particular, the result
of Ref.\cite{Amb-Eck} follows from this contribution if one assumes the
equilibrium Fermi-Dirac distribution and the multiplicity $p=2$. 
Indeed, using the identity $L\,\frac{\partial}{\partial \phi}\,G_{0}^{R,A}(E)=-
G_{0}^{R,A}(E)\hat{J}G_{0}^{R,A}(E)$ one concludes that ${\bf I_{2}}$ 
can be expressed in terms of the flux derivative of the effective energy
functional ${\cal E}_{{\rm eff}}(\phi)$ that in the absence of equilibrium stands
for
the total energy ${\cal E}(\phi)$ in Eq.(\ref{PC}). In order to understand the
physical meaning of the additional contribution ${\bf I_{1}}$ we invoke the basis
of exact {\it many-body} electron states $\Psi_{\mu}$ and the corresponding matrix
elements of the current operator $J_{\mu\nu}$. In this representation the 
`thermodynamic' contribution ${\bf I_{2}}$ is expressed in terms of the {\it
diagonal} matrix elements $J_{\mu\mu}$ only. In contrast to that the `kinetic'
contribution ${\bf I_{1}}$ contains only {\it off-diagonal} matrix elements
$J_{\mu\nu}$. One can  check that the   replacing the current operator
$\hat{J}$ in the expression for ${\bf I_{1}}$  by the unit operator 
results in vanishing of the whole expression, as it is required by the 
orthogonality of the {\it different} many-body wave functions. 
Thus the contribution ${\bf I_{1}}$ contains an information on the {\it overlap}
of the different many-body wave functions that is totally absent in  ${\bf I_{2}}$.

We also note that ${\bf I_{1}}$ comes entirely from the Fock-type diagrams,
while ${\bf I_{2}}$ contains both the Hartree and the Fock contributions. The 
two contributions have opposite signs and the balance between them is controlled by
the multiplicity factor $p$. Therefore the thermodynamic contribution ${\bf I_{2}}$
is very
sensitive to the parallel magnetic field that leads to the cancellation of the main
part of the persistent current which is due to the {\it real} part of the
screened interaction $U^{R,A}$. 
At the same time the effect of the parallel magnetic field on the kinetic
contribution reduces merely to a possible variation of the $E$-dependence of DoS.

The contribution $\langle{\bf I}_{1}\rangle$ averaged over disorder is exactly the
relaxation-induced non-equilibrium current $I^{(r)}(\phi)$. Substituting
Eq.(\ref{Uexp}) into the first of Eqs.(\ref{three}) 
we obtain 
\begin{equation}
\label{necurrent}
I^{(r)}_{n}=\int_{-\infty}^{+\infty}\frac{dE dE'}{(2\pi)^{2}}
\int_{-\infty}^{+\infty}d\omega\,
J_{n}(E,E',\omega)\,R_{\omega}(E,E'),
\end{equation}
where $R_{\omega}(E,E')$ is given by Eq.(\ref{R}) and $J_{n}(E,E',\omega)$ is the
$n$-th
flux harmonic of
\begin{equation}
\label{JJ}
-\sum_{q\neq 0}\langle
(G_{0}^{A}-G_{0}^{R})_{E-\omega}\,G_{0}^{A}(E)\hat{J}G_{0}^{R}(E)\rangle_{q}\;
U_{\omega,{\bf q}}(E').
\end{equation}
The disorder average $\langle ...\rangle$ in Eq.(\ref{JJ}) is done within the   
impurity diagrammatic technique. In the leading approximation in $1/g$
the result is described by the diagram Fig.2.
\begin{figure}[h]
\psfrag{k+q,0}{$\vec k+\vec q,0$}\psfrag{q,-w}{$\vec q,\omega$}\psfrag{k,-w}
{$\vec k,\omega$}
\psfrag{q,w}{$\vec q,\omega$}\psfrag{w}{$\omega$}\psfrag{-w}{$\omega$}\psfrag{0}{$0$}
\psfrag{R, E}{{\bf A},$E$}\psfrag{A, E}{{\bf R},$E$}\psfrag{R, E-w}{{\bf
A},$E-\omega$}
\includegraphics[height=3.5cm]{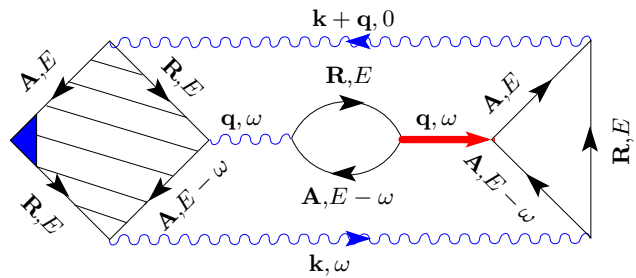}
\caption{Disorder averaging of the $A{\bf A}{\bf R}$ term in Eq.(\ref{JJ}). Solid
lines labelled by $R(A)$ are
disorder
averaged electron Green's functions $\langle G_{0}^{R(A)}(E)\rangle$ with the
energy indicated; the wavy lines are the diffusion propagators (diffusons and
cooperons) with the momentum and frequency indicated; the dashed square is the
Hikami box with
the vector vertex $\hat{J}$ denoted by the bold triangle; the bold solid line
is the screened electron interaction $ U_{\omega,{\bf q}}(E')$.}
\end{figure}
The quasi-one-dimensional geometry of rings is taken into account by the
quantization of momenta of diffusion propagators $[D_{E}{\bf q}^2
-i\omega]^{-1}$ and electron interaction $U_{\omega,{\bf q}}(E')$,
such that the transverse
momentum ${\bf q}_{\perp}=0$, and the
longitudinal momentum $q_{||}=
(2\pi/L) \,(m-2\phi/\phi_{0})$ for the cooperons and $q_{||}\equiv q=
(2\pi m/L)$ for the diffusons and the electron interaction, where $m=0,\pm
1,\pm2,...$ is an integer.
The only constraint is that the electron interaction must be zero at zero momentum.
Performing the Poisson summation over ${\bf k}$, we obtain
\begin{equation}
\label{J}
J_{n}(E,E',\omega)=
\frac{4e D_{E}}{g L^{2}}\,\Im \sum_{q\neq
0}\frac{(1-e^{-nL/L_{\omega}})(D_{0}^{2}q^{4}+\omega^{2}) }{(D_{E}q^{2}-i\omega)^{3}
(D_{E'}^{2}q^{4}+\omega^{2})},
\end{equation}
where $L_{\omega}=[D_{E}/(-i\omega)]^{\frac{1}{2}}$ and $g=\nu D S/L$.

The energy dependence of $J_{n}(E,E',\omega)$ originates i). from the 
$E$-dependence of the triangle of Green's functions in Eq.(\ref{JJ}) and ii). from
the $E'$-dependence of the polarization bubble in the effective interation
$U_{\omega, {\bf q}}(E')$. It is easy to see that only the latter is important.  
Indeed, let us expand $R_{\omega}(E,E')$ in Eq.(\ref{necurrent}) up to the linear in
$\omega$
term.  If
$D_{E'}=D_{0}$ one can perform the Wick
rotation $-i\omega \rightarrow
\omega$ 
that immediately gives
$\int_{0}^{\infty} J_{n}(E,E',\omega)\,\omega d\omega=0$ because of the $\Im$
sign in front of the sum in 
Eq.(\ref{J}). This is not true if
$\delta D_{E'}\neq 0$, as in this case the Wick rotation  leads to divergency.
A careful analysis shows that the term proportional to $\delta D_{E}$ in
Eq.(\ref{J}) makes a
contribution to Eq.(\ref{necurrent}) that is small by the parameter
$(E_{T}/T)^{1/2} \ll 1$ compared to that resulting from $\delta D_{E'}$.
Neglecting this contribution by setting $D_{E}=D_{0}$ we arrive at a finite result
Eq.(\ref{fres}).

Note that $J_{n}(E,E',\omega)$ is {\it not} exponentially small at
$\omega\gg E_{T}$ or $L\gg L_{\omega}$.
This can be traced back to  the structure 
$G_{0}^{R,A}(E-\omega)G_{0}^{{\bf R}}(E)\hat{J}G_{0}^{{\bf A}}(E)$ 
of the kinetic term ${\bf I_{1}}$ that
allows to build a cooperon at zero frequency \cite{KY1}. This is impossible for the 
thermodynamic term ${\bf I_{2}}$ in which both Green's functions with the same
energy $E$ are either retarded or advanced. Therefore the corresponding kernel 
for the thermodynamic term is proportional to $\exp[-L/L_{\omega}]\sim
\exp[-\sqrt{T/E_{T}}]$. This is the
reason why the kinetic term wins over the thermodynamic one for $T\gg E_{T}$. 

In conclusion, using the Keldysh formalism we identified two different
contributions, thermodynamic and kinetic,
to the disorder-averaged magnetization of mesoscopic rings with a non-equilibrium
distribution
of one-electron energies. Both contributions are caused by the electron-electron
interaction. However the kinetic contribution is present only out of equilibrium
provided that
the one-electron density of states is not constant near the Fermi energy.
This contribution is proportional to the same combination of the one-electron
energy distribution function as the inelastic relaxation rate and is thus
intimately related with the relaxation. 
In the basis of exact many-body wavefunctions of the weakly
interacting electron gas the kinetic contribution is strictly off-diagonal in
contrast to
the thermodynamic contribution (persistent current) that depends only on the
diagonal matrix elements of the
current operator. 
The sign of the kinetic contribution is not fixed by the basic symmetry of the
problem (orthogonal or symplectic) but depends on the nature of the energy
dependence of the one-electron DoS.

We are grateful to Igor Aleiner for critical comments and to B.L.Altshuler for
fruitful discussions clarifying the
physical meaning of the kinetic contribution to the current.

\end{document}